\documentclass[fleqn, usenatbib]{mnras}

\usepackage{graphicx}	
\usepackage{amsmath}	
\usepackage{amssymb}	
\usepackage{multicol}   
\usepackage[T1]{fontenc}
\usepackage{newtxtext, newtxmath}
\usepackage{xspace}

\newcommand\kms{$km\,s^{-1}$}
\newcommand\oucd{NGC\,936\_UCD\xspace}

\title[Formation of UCD through tidal threshing]{The creation of a massive UCD by tidal threshing from NGC\,936}

\author[Paudel et al.]{
Sanjaya Paudel,$^{1,2}$\,%
Pierre-Alain Duc,$^{3}$
Sungsoon Lim,$^{1}$
M\'elina Poulain,$^{4}$
Francine R. Marleau,$^{5}$ \newauthor
Oliver M\"uller,$^{6}$ 
Rubén Sánchez-Janssen,$^{7}$
Rebecca Habas,$^{3,8}$
Patrick R. Durrell,$^{9}$
Nick Heesters,$^{6}$\newauthor
Daya Nidhi Chhatkuli,$^{10}$
Suk-Jin Yoon,$^{1,2}$\thanks{E-mail: sjyoon0691@yonsei.ac.kr}
\\
$^{1}$Department of Astronomy, Yonsei University, Seoul, 03722, Republic of Korea\\
$^{2}$Center for Galaxy Evolution Research, Yonsei University, Seoul, 03722, Republic of Korea\\
$^{3}$Universit{\'e} de Strasbourg, CNRS, Observatoire astronomique de Strasbourg, UMR 7550, F-67000 Strasbourg, France\\
$^{4}$Space Physics and Astronomy Research Unit, University of Oulu, P.O. Box 3000, FI-90014, Oulu, Finland\\
$^{5}$Institut f{\"u}r Astro- und Teilchenphysik, Universit{\"a}t Innsbruck, Technikerstra{\ss}e 25/8, Innsbruck, A-6020, Austria\\
$^{6}$Institute of Physics, Laboratory of Astrophysics, École Polytechnique Fédérale de Lausanne (EPFL), 1290 Sauverny, Switzerland\\
$^{7}$UK Astronomy Technology Centre, Royal Observatory, Blackford Hill, Edinburgh, EH9 3HJ, UK\\
$^{8}$INAF Osservatorio Astronomico di Teramo, via Maggini, I-64100, Teramo, Italy\\
$^{9}$Youngstown State University, One University Plaza, Youngstown, OH 44555 USA\\
$^{10}$Central Department of Physics, Tribhuvan University, Kirtipur 44618, Kathmandu, Nepal
 }

\date{Accepted 2023 August 30. Received 2023 August 28; in original form 2023 August 03.}

\pubyear{2022}

\begin{document}
\label{firstpage}
\pagerange{\pageref{firstpage}--\pageref{lastpage}}
\maketitle

\begin{abstract}
We study a compact nucleus embedded in an early-type dwarf galaxy, MATLAS-167,  which is in the process of disruption by the tidal force of the neighboring giant S0 galaxy, NGC\,936, in a group environment. Using the imaging data of the MATLAS survey, we analyze the stellar tidal tail of MATLAS-167 and its central compact nucleus, designated as \oucd. We find that \oucd has a luminosity of M$_{g}$ = $-$11.43$\pm$0.01 mag and a size of 66.5$\pm$17 pc, sharing the global properties of Ultra Compact Dwarf galaxies (UCDs) but significantly larger and brighter compared to the typical UCD populations observed in the Virgo cluster. By integrating the total luminosity of both the tidal stream and MATLAS-167, we estimate that the disrupted dwarf progenitor possesses a luminosity of M$_{g}$ = $-$15.92$\pm$0.06 mag, a typical bright dE luminosity. With the help of the optical spectrum observed by the SDSS survey, we derive the simple stellar population properties of \oucd: a light-weighted age of 5.6$\pm$0.7 Gyr and metallicity of [Z/H] = $-$0.83$\pm$0.3 dex. Our findings suggest that tidal threshing is a possible formation mechanism of bright UCD populations in close proximity to giant galaxies.
\end{abstract}


\section{INTRODUCTION}

Ultra-compact dwarf galaxies (UCDs) bridge the gap between galaxies and star clusters in terms of mass, size, and luminosity, making it difficult to clearly distinguish between the two classes of stellar systems  \citep{hilker99,drinkwater00,Phillipps01,Evstigneeva08,Norris14}. The question at the heart of this discussion is whether UCDs are the largest star clusters or the smallest galaxies \citep{Mieske02,Patig06}. UCDs are larger, brighter, and more massive than the typical globular clusters (GCs) with typical half-light radii of 10 $\lesssim$ R$_{h}$ $\lesssim$ 100 pc,  and luminosities L$_{i}$ $\gtrsim$ 10$^{5}$ L$_{\sun}$ \citep{Hasegan05,Mieske08,Misgeld11,Norris14,Voggel16}. Their stellar population is old ($\gtrsim$5 Gyr), with a wide range of metal content, mostly sub-solar \citep{Firth09,Paudel10,Chilingarian11,Janz16,Zhang18,Forbes20,Fahrion19}. The central velocity dispersions ($\sigma_{v}$) of UCDs are similar to dwarf galaxies, with a typical value of 20 $\lesssim$ $\sigma_{v}$  $\lesssim$ 50 \kms. Their dynamical mass estimates show that they have mass-to-light ratios, which are, on average, about twice as large as those of GCs \citep{Hilker07,Baumgardt08,Frank11,Mieske13,Janz15}. Recent high spatial resolution spectroscopic observations show that a fraction of UCDs also hosts a central intermediate-mass black hole \citep{Seth14, Ahn17,Ahn18,Afanasiev18,Voggel19}.

\begin{figure*}
\includegraphics[width=16cm]{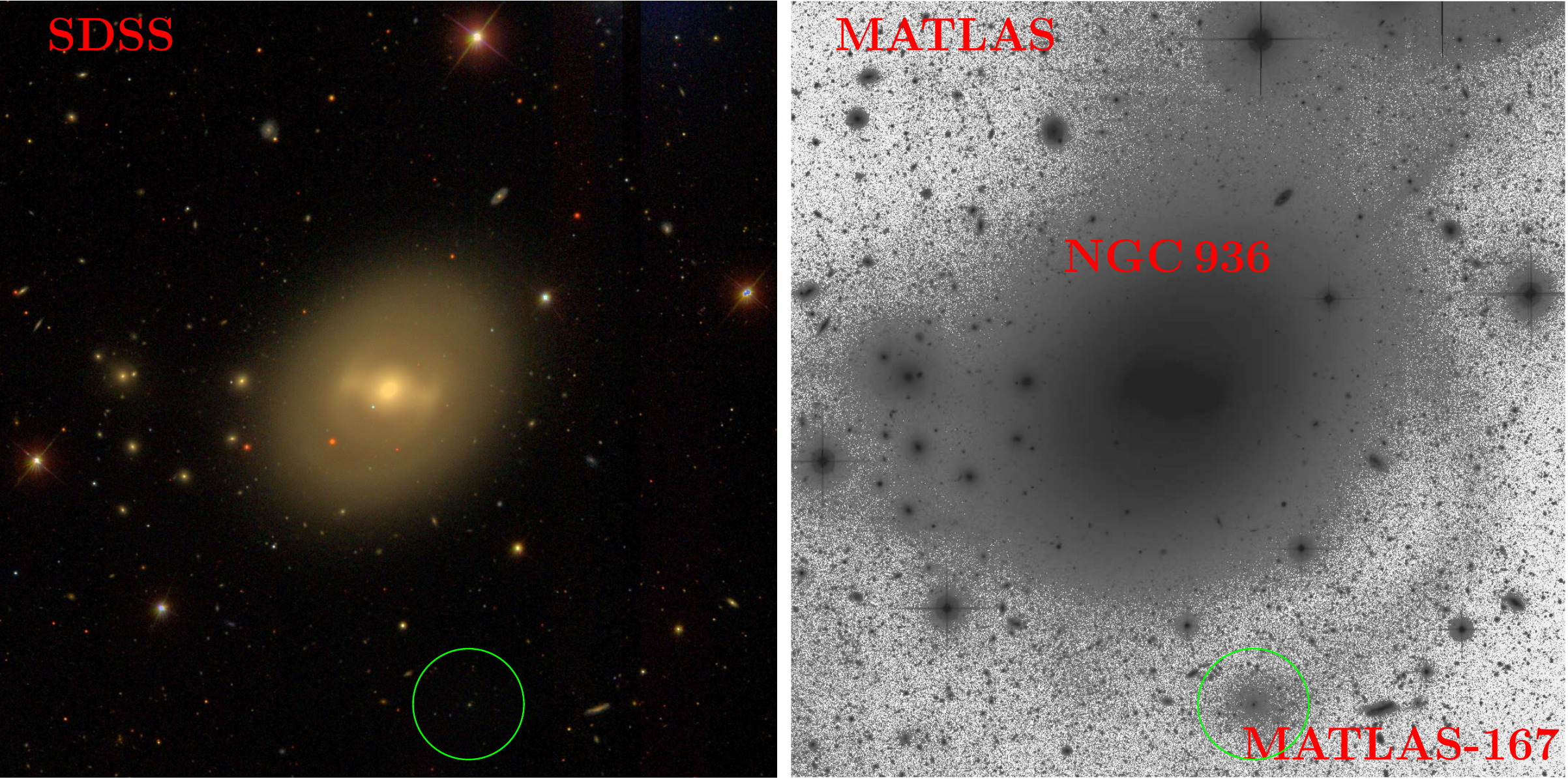}
\caption{Comparison between the SDSS and the MATLAS image. The left panel shows a tri-color image of NGC\,936 from SDSS, created by combining $g$-, $r$-, and $i$-band images. The right panel shows a deep $g$-band image from the MATLAS, which clearly reveals the filament and low surface brightness plumes around NGC\,936. Both images have a field of view of 5.5$\arcmin$$\times$5.5$\arcmin$. The position of the disrupted dwarf, MATLAS-167, is highlighted by a green circle in both images. While only the star cluster is visible in the SDSS image, the underlying low surface brightness host is revealed in the MATLAS image. }
\label{mainfig}
\end{figure*}

Since the discovery of UCDs, there has been a significant amount of research focused on understanding their origins. It has become clear that UCDs are not a uniform population and can be formed through a variety of different processes \citep{Hilker11}. Two main formation pathways are frequently discussed in the literature  \citep[e.g.,][]{Fellhauer02}. The first involves tidal disruption, with UCDs proposed as the remnant nuclei of tidally disrupted galaxies \citep{Drinkwater03,Gregg03,Goerdt08,Pfeffer13,Pfeffer14}. In this scenario, a nucleated dwarf galaxy in a cluster or group environment may undergo complete tidal disruption, leaving behind a naked dense stellar core (known as a nuclear star cluster). The remnant dense nuclear star cluster is gravitationally strong enough to retain its stars against tidal disruption \citep{Bekki03}. Evidence in support of the tidal disruption origin of UCDs includes the presence of features such as tidal tails, extended haloes, SMBH, and asymmetries around these objects \citep{Voggel16, Wittmann16,Schweizer18, Evstigneeva08,Liu20}. Other nucleated dwarf galaxies undergoing disruption have been discovered. They include the Sagittarius dwarf galaxy around the Milky Way, a so-called dog-leg tidal stream around NGC\,1407 \citep{Galianni10,Amorisco15}
and extremely diffuse nucleated dwarf galaxies at the Virgo cluster \citep{Mihos15}.

 The second scenario suggests that UCDs are the high-mass end of the GC mass function \citep{Kroupa98,Fellhauer02,Mieske02,Bruns11} and bright UCDs might have formed through the merger of GCs \citep{Patig06}. It is also argued that UCDs can be primordial objects formed in an intense burst of star formation \citep{Murray09}.

There is a wide range of properties among known UCDs, and they share characteristics with both GCs and the nuclei of dwarf galaxies. 
This suggests multiple formation processes contribute to their creation \citep{Francis12}. However, it is likely that stripped nuclei account for at least some percentage of the UCD population due to various similarities to compact galaxy nuclei \citep{Drinkwater03,Paudel10}. These include overlapping luminosity distributions and similar size$-$luminosity relationships \citep{Evstigneeva08}, internal velocity dispersions \citep{Drinkwater03}, positions on the color$-$magnitude diagram, and stellar population properties \citep{Cote06,Evstigneeva08,Paudel10,Brodie11,Chilingarian11,Spengler17,Zhang18}.

In this work, we identify a star cluster located at the end of a tidal stream that is likely to have originated from the disruption of an early-type dwarf galaxy (dE), MATLAS-167. The star cluster is bright, $M_{g}$ = $-$11.43$\pm$0.01 mag, and compact, likely a surviving nucleus of MATLAS-167 disrupted by the tidal force of the nearby giant galaxy NGC\,936 located at 23.0 Mpc\footnote{The distance is measured using the surface brightness fluctuation method by \cite{Tonry01}.} away from us. We propose that the nuclear star cluster is in the process of forming a UCD  through tidal stripping. 

\section{Data and Analysis}
The aim of the Mass Assembly of early Type gaLAxies with their fine Structures (MATLAS) project is to conduct a comprehensive imaging survey of local elliptical galaxies that were selected from the ATLAS$^{3D}$ legacy survey \citep{Cappellari11,Duc15}. Its primary objective is identifying and documenting low surface brightness features such as stellar streams, filaments, and shells surrounding giant early-type and dwarf galaxies \citep{Bilek20,Habas20,Marleau21}. This project has a magnitude limit of 29 mag arcsec$^{2}$ for extended low surface brightness objects.  Through a thorough visual examination of all the galaxies in the survey, we have identified a system of ongoing disruption of a dwarf galaxy around the giant S0 galaxy, NGC\,936. The disrupting dwarf galaxy is MATLAS-167, and it is cataloged as a dE galaxy in the dwarf galaxy catalog, which has a prominent bright nucleus at the center \citep{Poulain21}.

In Figure \ref{mainfig}, we compare the SDSS color image and the MATLAS $g$-band image. As expected, the SDSS image does not reveal any stream, and only a compact source is visible (see the green circle). On the other hand, the deeper MATLAS $g$-band image displays a spectacular view of the tidal stream around  NGC\,936. The compact star cluster is embedded in a stellar stream, which we have marked by a green circle. We consider it a putative UCD (hereafter \oucd). It is located at the end of the stream, which forms an almost semi-circular trajectory around NGC\,936. The focus of this study is the nature of the interaction between NGC\,936 and MATLAS-167 and the evolution of \oucd.

\begin{figure}
\includegraphics[width=8.5cm]{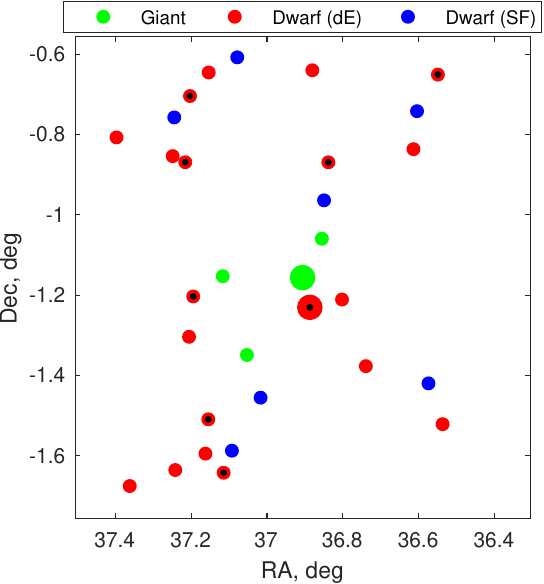}
\caption{On-sky position of member galaxies in the NGC\,936 group. Green symbols represent giant galaxies, while dEs and star-forming dwarf galaxies are represented by red and blue symbols, respectively. Black dots indicate nucleated dEs. Additionally, the diagram includes two large symbols to indicate the position of NGC\,936 itself and its disrupted satellite, MATLAS-167. }
\label{group}
\end{figure}

NGC\,936 is a barred S0 galaxy classified as S0Bb in the RC3 catalog \citep{Vaucouleurs91}. It is the most dominant galaxy in the group, which includes three other massive galaxies. It has a face-on orientation with an inclination of $<$10 degrees as shown in Figure \ref{mainfig}, and it has a prominent central bar. The MATLAS search for dwarf galaxies identified 27 dwarf galaxies around NGC\,936, and their distribution around NGC\,936 is shown in Figure  \ref{group}. Only 7 out of 27 dwarf galaxies are star-forming dwarf galaxies \citep{Habas20,Poulain21}. Among 20 dEs, the nucleated fraction is nearly 50\%.

 \begin{table}
 \begin{center}
\caption{Physical properties of \oucd}
\begin{tabular}{cccc}
\hline
Properties & Values & Unit & Note\\
\hline
R.A. & 02:27:32.88 & h:m:s & 1\\
Decl. & $-$01:13:49.31 & d:m:s &  2\\
$M_{g}$ & $-$11.43$\pm$0.01 & mag & 3 \\
$z$ & 0.0039 & & 4\\
$g-r$  & 0.60$\pm$0.01 & mag & 5\\
R$_{e}$ & 66.5$\pm$17 & pc& 6\\
$M_{g}$ &  $-$15.92$\pm$0.06 & mag& 7\\
$\Delta$d &  23 & kpc&8\\
$\Delta$v$_{r}$ & 260 & \kms & 9\\
\hline
\end{tabular}
\end{center}
1) R.A. of \oucd \\
2) Decl. of \oucd \\
3) The absolute $g$-band magnitude of \oucd \\
4) Redshift measured from the SDSS spectrum of \oucd \\
5) $g-r$ color of \oucd \\
6) Effective radius of \oucd \\
7) The $g$-band integrated magnitude of putative dwarf galaxies along the streams\\
8) Sky-projected separation between \oucd\ and NGC\,936\\
9) Relative line-of-sight velocity between \oucd\ and NGC\,936
\label{tab1}
\end{table}

\subsection{Imaging and Photometry}
\subsubsection{The nucleus}
In this work, we used Megacam CHFT images obtained by the MATLAS survey \citep{Duc15,Bilek20}. The MATLAS survey consisted of $g$, $r$, and $i$-band images, where the $g$-band is the deepest and $i$-band is the best in image quality. We, therefore, used the $g$-band images for the photometric measurement and surface photometry of the host galaxy. The $i$-band, particularly, was used for size measurement of compact nucleus, which can provide a better spatial resolution than others. All $g$, $r$, and $i$-band images are observed in 0.19$\arcsec$ pixel$^{-1}$ spatial resolution, and the $i$-band has a median PSF of 0.89$\arcsec$  which corresponds to
99 pc  at the distance of NGC 936 (23 Mpc).

 \begin{figure}
\includegraphics[width=8cm]{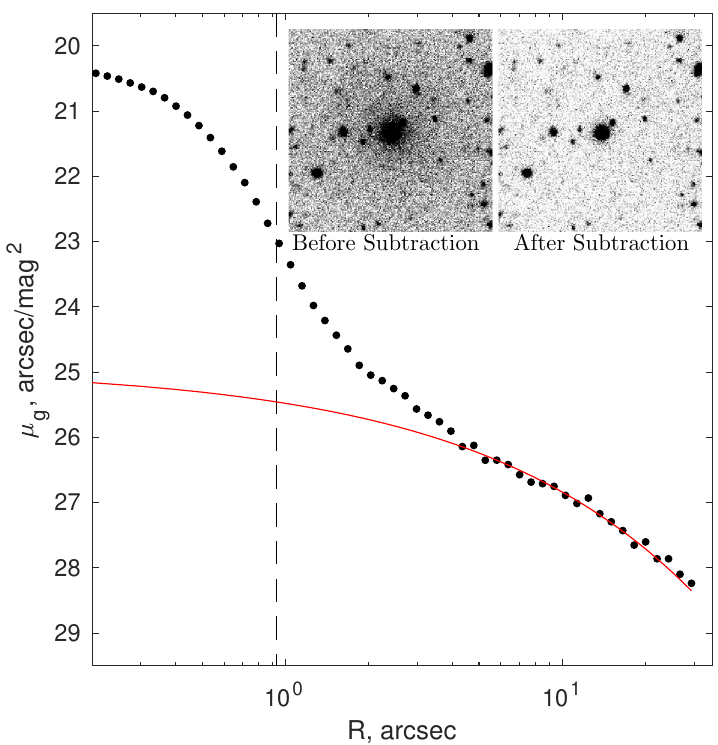}
\caption{The $g$-band surface brightness profile of MATLAS-167 along its major axis. The best-fit S\'ersic function is shown by the red line. We also show a 45$\arcsec$$\times$45$\arcsec$ $g$-band image of MATLAS-167  and the residual after subtracting the best-fit model image in the inset. The vertical dash line represents the size of \oucd.}
\label{profit}
\end{figure}
To accurately measure the flux of the nucleus, ensuring that the surrounding galaxy light does not contaminate it, we employed a method that involves subtracting the host galaxy light. To accomplish this, we utilized the IRAF $ellipse$ task, which outputs an azimuthally averaged value along an elliptic path with a function of galactocentric radii. Figure  \ref{profit} depicts the $g$-band light profile of MATLAS-167 along the major axis, with the black dots representing the observed data points and the red line representing the best-fitted S\'ersic function. To avoid any interference from the central nucleus, we excluded the inner  (r $\le$ 4$\arcsec$) data points during the fit.

The best-fitted parameters derived from the best-fitted S\'ersic function are an effective radius (R$_{e}$) of 12.23$\arcsec$ and a S\'ersic index (n) of 1.4. To construct a two-dimensional representation of the observed galaxy, we incorporated the one-dimensional best-fitted flux into the output of the IRAF ellipse fit and employed the $bmodel$ task. Considering this best-fit S\'ersic model represents the bound component of MATLAS-167, it has a luminosity of $M_{g}$ = $-$12.76 mag.

Subsequently, we performed aperture photometry of the compact nucleus in the resulting model-subtracted residual images. For the measurement of the total flux and its magnitude, we used an aperture that is roughly twice the size of the Full Width at Half Maximum (FWHM). To determine the FWHM, we utilized multiple bright, unsaturated stars in the field as references. To eliminate background contributions, we selected an annulus with inner and outer radii of twice and thrice the FWHM, respectively. Total brightness is $M_{g}$ = $-$11.43$\pm$0.01 mag, and $g-r$ color is 0.6$\pm$0.01 mag.

\begin{figure}
\includegraphics[width=8.5cm]{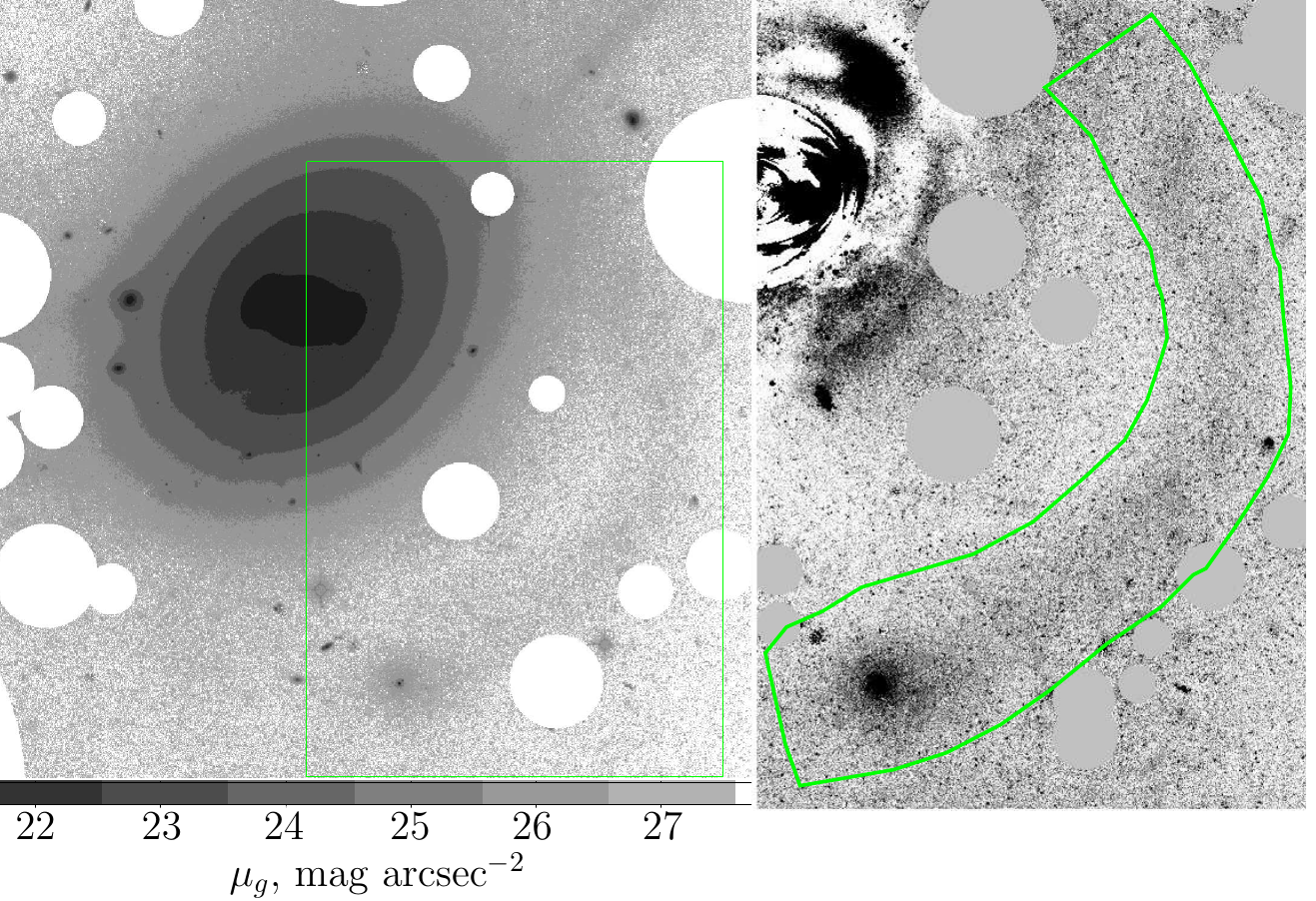}
\caption{The $g$-band surface brightness map of the field around NGC\,936. The unrelated foreground and background objects are masked out manually. The green box in the left panel represents the zoom-in area shown in the left panel, which is prepared after subtracting the model of NGC
936. The green polygon in the left panel delineates the aperture used to
carry out the photometry.} 
\label{magmap}
\end{figure}

\begin{figure}
\includegraphics[width=8cm]{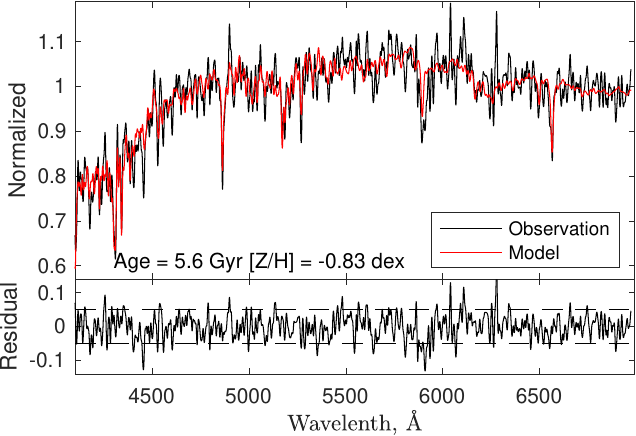}
\caption{The SDSS fiber spectrum (black), together with its best-fit SSP model spectrum (red).  The residuals are shown in the lower panel. The fit is generally consistent within 5 percent of the observed flux (the horizontal lines).}
\label{spect}
\end{figure}

The size of \oucd\ was determined by analyzing the galaxy-subtracted $i$-band image, where it was partially resolved. To perform the measurement, we utilized the publicly available software $ishape$, and explored both MOFFAT and KING profiles \citep{Larsen99}. The software convolves a model light profile with a provided PSF and fits it to the source. The analysis resulted in a size estimate of 0.56$\arcsec$ for the KING15 profile and 0.64$\arcsec$ for the MOFFAT15 profile, both exhibiting a similar uncertainty of 0.16$\arcsec$. When translated into physical units, these values correspond to sizes of 62 pc and 71 pc for the KING15 and MOFFAT15 profiles, respectively. The discrepancies in residuals between the two models were not statistically significant. Consequently, we opted to adopt the average of these two measurements, yielding a final size estimate of 66.5$\pm$17 pc. This is relatively large for a typical NSC of stellar mass $<$10$^{7}$ M$_{\sun}$ and NSC of the stellar mass of $\approx$10$^{7}$ M$_{\sun}$ or $M_{g}$$\approx$$-$12 mag typically have effective radius of $\approx$50 pc \citep{Boker04,Georgiev16}

\subsubsection{Surface photometry of the tidal stream}
A ring filter was utilized to remove foreground stars and compact background galaxies from the images, and any residual artifacts were manually subtracted using the IRAF task $imedit$. In Figure \ref{magmap}, we show the $g$-band surface brightness map of the field around NGC\,936 after cleaning and masking unrelated foreground and background objects. The background gradient of halo light from the nearby giant galaxy NGC\,936 is subtracted. First, a constant sky-background level is subtracted across the entire image. The constant sky background level is derived using 10 independent sky regions of size 10$\times$10 pixel boxes from which we sampled the sky background and calculated an overall median. Subsequently, we masked MATLAS-167 and its tidal tail region and ran $ellipse$ task to model NGC\,936, and then subtracted this model of NGC NGC\,936 from the image.

To measure the total flux of filamentary structure, we conducted aperture photometry using a polygonal aperture (see the green polygon in Figure \ref{magmap}). Since the surface brightness of the faint filaments was too low for automatic detection, an aperture is defined visually.  We excluded pixels below the S/N threshold from the measurement, and the resulting values are presented in Table \ref{tab1}. The total brightness in $g$-band we measured was M$_{g}$ = $-$15.92$\pm$0.04 mag. However, we want to emphasize that this estimate may not account for additional starlight below our detection threshold or behind NGC 936, and it may also include contamination from faint point sources. Therefore, caution should be exercised when interpreting these measurements as the accreted galaxy luminosity. We followed a similar procedure in the $r$-band image, measuring the flux within the identical polygonal aperture, and found that the color of the full stream is $g-r$ = 0.72$\pm$0.06 mag.

\subsection{Spectroscopy}
The SDSS targeted  \oucd for spectroscopic observation, which we retrieved from the SDSS archive server, and it proved to be of sufficient quality and high signal-to-noise ratio to perform a detailed stellar population study. The SDSS spectrum is observed with a fiber of radius 1.5$\arcsec$, which is nearly three times \oucd size. However, the light contribution of \oucd in the fiber is dominant, i.e., $>$90\%.

To extract the maximum amount of information from the spectrum, we employed a full-spectrum fitting method, which exploits the extensive wavelength coverage of SDSS optical spectroscopy. This fitting method involves modeling the spectrum using a combination of simple stellar populations (SSPs) defined by their age and metallicity. We utilized the publicly available code UlySS by \cite{Koleva08} for this purpose. We used an SSP model provided by \cite{Vazdekis10}, based on MILES stellar library \citep{Blzquez06}. This model considers the effects of different stellar evolutionary phases, such as the main sequence, red giant branch, and asymptotic giant branch. We fitted the observed spectrum of wavelength range 4100 to 7000 \AA\ after smoothing the SDSS spectrum by a three-pixel Gaussian kernel.  The quality of the model comparison with the SDSS spectrum is shown in Figure \ref{spect}, where the observed spectrum typically matches within 5 percent of the modeled flux. The analysis yielded a light-weighted SSP age of 5.6$\pm$0.8 Gyr and [Z/H] of $-$0.83$\pm$0.3 dex.

\section{Discussion}
\begin{figure}
\includegraphics[width=8cm]{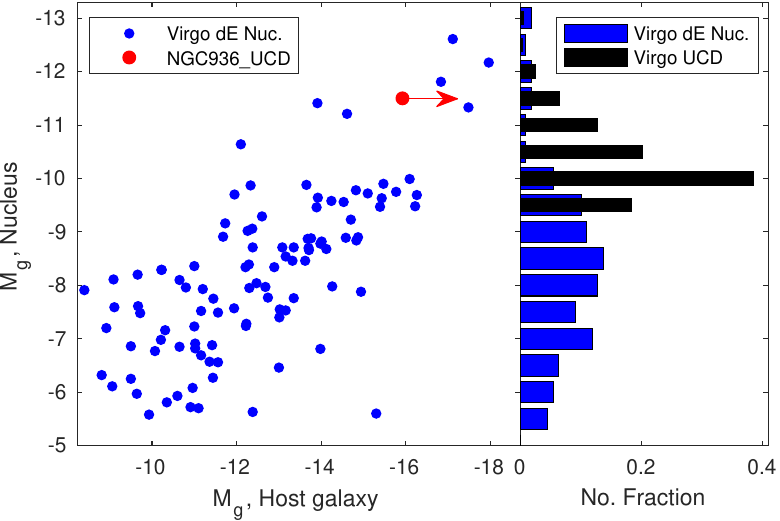}
\caption{Relation between the luminosity of dEs and their nuclei. \oucd is shown in red and the comparison sample of the Virgo cluster dEs are shown in blue, which we obtained from \citet{Sanchez19}. The arrow in \oucd represents our measurement of MATLAS-167 flux is a lower limit.}  the  In the right panel, we show the magnitude distribution of the Virgo cluster dE nuclei and UCDs. The Virgo cluster UCDs magnitudes are obtained from \citet{Liu20}. 
\label{nufrac}
\end{figure}

\subsection{Comparison of UCDs and dE Nuclei Properties}


UCDs and dE nuclei are compact and dense stellar systems of high mass. They often contain predominantly old stellar populations, indicating their formation in the early stages of galaxy evolution. In this section, we make a comparative analysis between UCDs and dEs nuclei and find the position of \oucd. For this purpose, we use the Virgo cluster UCDs and dE nuclei as reference samples.

The relationship between the luminosity of dEs and their nuclei is depicted in Figure  \ref{nufrac}. The Virgo cluster dE sample is obtained from \cite{Sanchez19}, shown in blue. \oucd is represented by a red dot. As anticipated, a well-established correlation emerges between the luminosity of dEs and that of their nuclei, placing \oucd among the brightest objects situated in the upper-right corner. It is important to note, however, that the estimated luminosity of the \oucd host galaxy represents a lower limit, implying that its actual position on the plot may have been even further to the right.

\begin{figure}
\includegraphics[width=8cm]{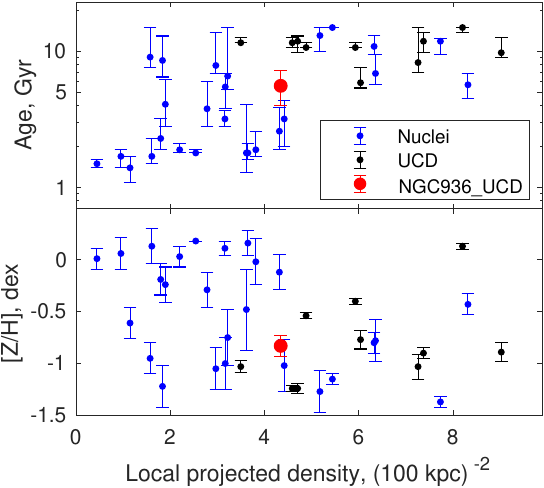}
\caption{Comparison of age and metallicity of the Virgo cluster UCDs (black) and dEs nuclei (blue) with respect to the local projected density. The data are from \citet{Paudel10}. \oucd is shown in red.}
\label{ssp}
\end{figure}

Figure \ref{ssp} illustrates the relationship between the derived SSP properties and local projected density. In this analysis, we utilized UCD and dE nuclei samples from the Virgo cluster, as studied by \cite{Paudel10}. The local density was determined by calculating the circular projected area enclosing the 10th neighbor. The results indicate a weak correlation between the local projected density and the ages of the nuclei. More importantly, an age break is observed at approximately $\sim$\,4 (100 kpc)$^{2}$. Almost all UCDs are located in the high-density region as defined above, and their age distribution overlaps with that of dE nuclei situated in high-density environments. A similar trend is identified in the metallicity distribution of dE nuclei, where those in high-density environments exhibit lower metallicity compared to nuclei of dE located in low-density environments. Notably, the SSP properties of \oucd, being situated in a relatively dense region, $>$4 (100 kpc)$^{2}$, resemble those of Virgo UCDs or dE nuclei located in dense regions.

\begin{figure}
\includegraphics[width=8cm]{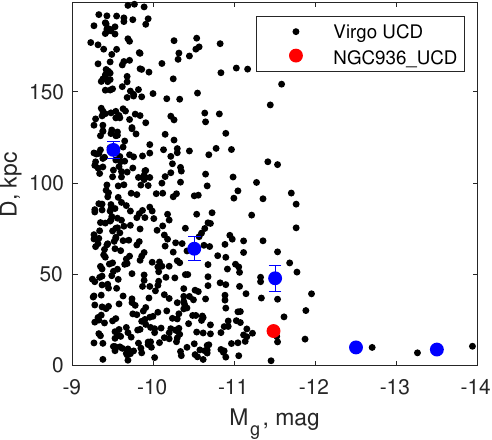}
\caption{Relation between the distance of UCDs from their nearest bright galaxy and their luminosities. The blue symbol represents the median distance in the magnitude bin, accompanied by an error bar that indicates the normalized standard deviation. \oucd\ is shown in red. }
\label{magdis}
\end{figure}

\subsection{ Tidal Interaction and Formation of UCDs}

Observations have shown that UCDs have a size$-$luminosity distribution and internal velocity dispersion similar to compact nuclei \citep{Drinkwater03,Evstigneeva08,Pfeffer13}. The high dynamical mass-to-light ratios of UCDs suggest that they may contain a significant amount of dark matter, which may have been inherited from the parent dwarf galaxies during the tidal disruption \citep{Baumgardt08,Mieske08}. Several UCDs display indications of asymmetrical or tidal features, while others reveal the presence of stellar envelopes or the status of transitional objects from dwarf galaxies to UCDs \citep{Wittmann16}.

State-of-the-art high-resolution imaging and spectroscopic observations of these compact objects have allowed us to search for the presence of supermassive black holes (SMBH). Particularly, recent observations have revealed that all the top three most massive UCDs of the Virgo cluster possess SMBH \citep{Ahn17,Ahn18,Seth14} and these SMBHs account for a substantial portion of their overall mass. These trends provide compelling evidence of their tidal stripping origin \citep{Voggel19}.

The tidal threshing scenario has been proposed to account for the origin of intra-cluster GCs, which is quite faint compared to the UCDs \citep{West95}. Our analysis suggests that massive UCDs are likely to form through tidal stripping. Based on Figure \ref{nufrac}, it is evident that the disrupted nucleated dE, MATLAS-167, stands out as one of the brighter dEs, and its nucleus luminosity is comparable to the brighter UCDs observed in the Virgo cluster. In fact, considering the combined luminosity of MATLAS-167 and its tidal stream, it surpasses the luminosity of all other dEs identified by the MATLAS dwarf galaxy survey around NGC\,936 \citep{Habas20}.

The close proximity of  \oucd to a giant galaxy raises questions about whether the special environment plays a role in the formation and evolution of bright UCDs. To shed light on this issue, we show a relation between UCD brightness and distance to the nearest bright galaxy (M$_{r}$ $<$$-$19 mag) of the Virgo cluster UCD sample studied by \citet{Liu20} in Figure \ref{magdis}. The figure reveals that bright UCDs tend to be closer to bright galaxies than faint UCDs, indicating a potential link between bright UCD formation and proximity to a bright galaxy. We find that almost all UCDs of $M_{g}$ < $-$12 mag are located within 20 kpc sky-projected distance from their nearest giant neighbor galaxy. \oucd, located at a sky-projected distance of 19 kpc away from a giant galaxy, NGC\,936, is consistent with the observed trend in the Virgo cluster. To quantify the observed trend, we sub-sample the UCD sample into faint (M$_{g}$ $>$ $-$11 mag) and bright categories and compute the two-point correlation coefficient between these subsets and massive galaxies. We find a significant disparity in the correlation coefficients. Specifically, the correlation coefficient between bright galaxies and bright UCDs is almost double (2.05) compared to that of bright galaxies and faint UCDs (0.96).

Indeed, the destruction of a bright dwarf necessitates a strong tidal force, which can typically be attained in the vicinity of a massive galaxy or within a densely populated cluster core. Consequently, the substantial tidal force exerted by giant galaxies appears to be advantageous in destroying bright dwarf galaxies, thereby resulting in exposed luminous nuclei commonly referred to as UCDs. This line of reasoning strongly supports the hypothesis that the tidal stripping mechanism is not only accountable for the formation of low-mass intra-cluster GC but also massive UCDs. Remarkably, these objects represent the two extremes of the mass function of compact stellar systems.

\section*{Acknowledgements}
S.-J.Y. and S.P. acknowledge support from the Basic Science Research Program (2022R1A6A1A03053472) through the National Research Foundation (NRF) of Korea. S.P. and S.-J.Y., respectively, acknowledge support from the Mid-career Researcher Program (No. RS-2023-00208957) and the Mid-career Researcher Program (No. 2019R1A2C3006242) through the NRF of Korea. O.M. is grateful to the Swiss National Science Foundation for financial support under the grant number PZ00P2\_202104. M\'elina Poulain is supported by the Academy of Finland grant n:o 347089.

\section*{DATA AVAILABILITY}
Most of the data underlying this article are publicly available.
The derived data generated in this research will also be shared on reasonable request to the corresponding author.




\bsp	
\label{lastpage}
\end{document}